\documentclass[notitlepage, superscriptaddress, showpacs, nobalancelastpage, twocolumn, aps, prl]{revtex4-1}

\usepackage[english]{babel}
\RequirePackage[T1]{fontenc}

\usepackage{siunitx} 
\usepackage[italicdiff]{physics}
\usepackage{amsfonts}
\usepackage{subfigure}
\usepackage{amsmath}
\usepackage{amssymb}
\usepackage{graphicx, epstopdf}
\usepackage{appendix}
\usepackage{xspace}
\usepackage{lipsum}
\usepackage{array}
\usepackage{hyperref}
\usepackage{xcolor}
\usepackage{my_symbols}
\usepackage{CJK}
\usepackage{bm}
\usepackage{verbatim}
\usepackage{tikz}
\usepackage[normalem]{ulem}

{\par}

\newcounter{mycomment}

\newcommand\rmv{\bgroup\markoverwith {\textcolor{red}{\rule[0.5ex]{2pt}{0.4pt}}}\ULon}

\newcommand\id{d}

\begin{document}
\begin{CJK*}{UTF8}{gbsn} 

\title{Skew scattering and side jump of spin wave across magnetic texture}

\author{Jin Lan (兰金)}
\affiliation{Center for Joint Quantum Studies and Department of Physics, School of Science, Tianjin University, 92 Weijin Road, Tianjin 300072, China}
\affiliation{Department of Physics and State Key Laboratory of Surface Physics, Fudan University, Shanghai 200433, China}


\author{Jiang Xiao (萧江)}
\email[Corresponding author:~]{xiaojiang@fudan.edu.cn}
\affiliation{Department of Physics and State Key Laboratory of Surface Physics, Fudan University, Shanghai 200433, China}
\affiliation{Institute for Nanoelectronics Devices and Quantum Computing, Fudan University, Shanghai 200433, China}

\begin{abstract}
Spin wave and magnetic texture are two elementary excitations in magnetic systems, and their interaction leads to rich magnetic phenomena.
By describing the spin wave and the magnetic texture using their own collective coordinates,
we find that they interact as classical particles traveling in mutual electromagnetic fields.
Based on this unified collective coordinate model, we find that both skew scattering and side jump may occur as spin wave passing through magnetic textures.
The skew scattering is associated with the magnetic topology of the texture, while the side jump is correlated to the total magnetization of the texture.
We illustrate the concepts of skew scattering and side jump by investigating the spin wave trajectories across the topological magnetic Skyrmion and the topologically trivial magnetic bubble respectively.
\end{abstract}

\maketitle
\end{CJK*}

Spin wave is the propagating excitation in ordered magnetizations, and magnetic texture is the stable collection of inhomogeneous magnetizations.
The interplay between spin wave and magnetic texture is a topic of long-standing interest in magnetism \cite{kruglyak_magnonics_2010, chumak_magnon_2015}.
The attention to it stems from the rich fundamental physics as well as its intimate connection to technological applications.
Using magnetic texture and spin wave for information storage and processing respectively, and their interaction for local information communication, a pure magnetic computing scheme can be realized \cite{han_mutual_2019, yu_magnetic_2020}.

Generally, the spin wave scattering by magnetic texture can be investigated by linearizing the Landau-Lifshitz-Gilbert equation into a Schrodinger-like \cite{yan_all-magnonic_2011, kovalev_thermomagnonic_2012-1, van_hoogdalem_magnetic_2013-1, lan_spin-wave_2015, yu_magnetic_2016, tatara_effective_2019}
or Klein-Gordon-like \cite{tveten_antiferromagnetic_2014, shiino_antiferromagnetic_2016, yu_polarization-selective_2018, qaiumzadeh_controlling_2018-1, kim_tunable_2019} equations, for which the influences of the inhomogeneous magnetic texture is included as effective electromagnetic fields \cite{dugaev_berry_2005, van_hoogdalem_magnetic_2013-1, tatara_effective_2019, kim_tunable_2019}.
Alternatively, the spin wave scattering are studied using Gaussian beam \cite{gruszecki_influence_2015, yu_magnetic_2016, wang_goos-hanchen_2019}, with the scattering details tracked by corresponding trajectories.
In the meantime, the dynamics of magnetic texture can be reduced to evolution of several collective coordinates \cite{thiele_steady-state_1973, schryer_motion_1974, thiaville_micromagnetic_2005, tretiakov_dynamics_2008}.
However, systematic and quantitative modeling of the interplay between spin wave and magnetic texture is still lacking.

In this work, we use collective coordinates to formulate the dynamics of spin wave and magnetic texture simultaneously in a Lagrangian frame. By analyzing the trajectories in the corresponding parametric spaces, we show that spin wave may experience both skew scattering and side jump as passing through magnetic texture.
As two basic deflections patterns,  skew scattering and side jump complete the scenario for spin wave penetrating the fictitious electromagnetic fields induced by magnetic texture.

\emph{Model.} Consider a magnetic system with its magnetization denoted by unit vector $\mb$, then the system Lagrangian is
\begin{equation}
\label{eqn:Lag}
 L = \int \cL \dd{\cV}
 =\int \qty[\bm{\Lambda}\cdot \dot{\mb} -u(\mb) ] \dd{\cV},
\end{equation}
where $\dot{\mb}\equiv \partial_t \mb$, 
$u(\mb)$ is the magnetic energy density, and $\cV$ is the system volume. The first term in \Eq{eqn:Lag} is the Wess-Zumino action for the spin precession dynamics, where $ \bm{\Lambda}(\bm{\Omega}) = (\bm{\Omega}\times\mb )/ (1+\mb\cdot\bm{\Omega})$ is the vector potential of a magnetic monopole with arbitrary direction $\bm{\Omega}$ \cite{haldane_geometrical_1986}. The dissipation of magnetic system is included by the Rayleigh function accompanying \Eq{eqn:Lag}: $R = \int (\alpha/2) (\dot{\mb}\cdot \dot{\mb}) \dd{\cV}$ with $\alpha$ the Gilbert damping constant. The magnetic dynamics corresponding to \Eq{eqn:Lag} is the Landau-Lifshitz-Gilbert (LLG) equation.

The magnetic energy density $u(\mb)$ generally reads
\begin{equation}
\label{eqn:mag_energy}
u(\mb) 
= \frac{K}{2} \qty[1-(\mb\cdot \hbz)^2]
+ \frac{A}{2} (\nabla \mb)^2 +\frac{D}{2} \mb\cdot(\nabla \times \mb),
\end{equation}
where $K$ is the easy-axis anisotropy along $\hbz$, $A$ is exchange coupling constant, and $D$ is (bulk-type) Dzyaloshinskii-Moriya interaction (DMI) constant.
This energy density is minimized with $u = 0$ for a homogeneous magnetic domain with magnetization pointing along the easy-axis $\mb_0 = \pm \hbz$. Depending on the exact values of parameter $K, A, D$, it also supports inhomogeneous magnetic textures such as the magnetic domain wall, Skyrmion \cite{rosler_spontaneous_2006, muhlbauer_skyrmion_2009-1, yu_real-space_2010}, or  bubble \cite{koshibae_berry_2016, phatak_nanoscale_2016, loudon_images_2019}, \etc.
Here we mainly focus on the magnetic system where exchange coupling is dominating,
therefore the long range dipolar interaction is ignored in \Eq{eqn:mag_energy}.

The dynamical magnetization $\mb$ can be divided into the slowly-moving texture background $\mb_0$ and the fast-evolving spin wave excitation part $\mb'$, \ie $\mb = \mb_0+\mb'$.
Due to the unity condition $|\mb|=1$, the excitation part satisfies the transverse condition: $\mb_0 \cdot \mb' = 0$ everywhere.
By defining the local spherical coordinates $\hbe_{r, \theta, \phi}$ with $\hbe_r = \mb_0$ and $\hbe_{\theta, \phi}$ as two transverse directions of $\mb_0$, the spin wave is expressed as $\mb'=m_\theta\hbe_\theta+m_\phi\hbe_\phi$, or equivalently as a complex field $\psi(\br, t) =m_\theta -i m_\phi$.

We proceed by dividing the Lagrangian density in \Eq{eqn:Lag} into the contributions from the texture background $\mb_0$, the excitation part $\psi$,
and their interactions: $\cL = \cL_0 + \cL' + \cL_\ssf{int}$.
The Lagrangian for magnetic texture is $\cL_0= a^0_t -u_0$, where $a^0_t= \bm{\Lambda}_0 \cdot \dot{\mb}_0$ is for texture kinetics,  $u_0\equiv u(\mb_0)$ is the local texture energy associated with the texture background $\mb_0$. The Lagrangian for spin wave is $\cL' = (i \psi^* \dot{\psi} - A \nabla\psi^* \cdot \nabla \psi -K \psi^*\psi)/2$.
When spin wave travels upon magnetic texture, their interaction is described by
\begin{equation}
 \label{eqn:Lag_int}
 \cL_\ssf{int} 
 = -  \bj \cdot \ba - \varrho \qty(  a_t^0 - 2 u_0),
\end{equation}
where $\bj= -i A(\psi^* \nabla \psi -\psi \nabla \psi^*)/2$ is the spin wave flux,
$\varrho= \psi^*\psi/2$ is the local intensity.
Here $a^0_t$ coincides with the geometrical scalar potential due to basis variation $\hbe_{\theta/\phi}$,
when specifically choosing $\bm{\Lambda}_0 \to [\bm{\Lambda}_0(+\hbz) + \bm{\Lambda}_0(-\hbz)]/2$.
In addition, $\ba= \ba_0+\ba_\ssf{D}$  is the total vector potential,
including the contribution from the inhomogeneous magnetization: $\ba_0= \bm{\Lambda}_0 \cdot \nabla\mb_0$ \cite{dugaev_berry_2005, van_hoogdalem_magnetic_2013-1, tatara_effective_2019, kim_tunable_2019} and the contribution from the DMI: $\ba_\ssf{D} = (D/2A) \mb_0$ \cite{van_hoogdalem_magnetic_2013-1, kim_tunable_2019}.
In addition, there is an energy reduction $2\varrho u_0$ to texture energy in \Eq{eqn:Lag_int}, since the magnitude of the magnetization reduces to $\mb_0 \sqrt{1-(\mb'\cdot \mb')}$ due to spin wave excitation $\mb'$.
Due to this energy reduction $2\varrho u_0$,  spin wave is energetically more favorable inside magnetic texture, leading to spin wave bound state \cite{lan_spin-wave_2015, garcia-sanchez_narrow_2015, wagner_magnetic_2016, sluka_emission_2019}.

\emph{Magnetic dynamics in collective coordinates.} The magnetic texture is relatively fixed in its shape, thus its dynamics can be captured by a series of collective coordinates $\qty{X_\mu(t)}$ with $\mu = 1, 2, 3, \dots$, \ie $\mb_0(t) \equiv \mb_0[\qty{X_\mu(t)}]$.
In the meantime, we assume the spin wave excitation is in the form of wave packets, which also has a fixed shape and can be described by its central position $\bx=\qty{x_i(t)}$ with $i=1, 2, 3$, \ie $\psi(t)\equiv \psi[\bx(t)]$.
In terms of these collective coordinates for slowly varying texture \cite{thiele_steady-state_1973, thiaville_micromagnetic_2005, tretiakov_dynamics_2008, tveten_staggered_2013-2} and the generalized coordinates for the fast varying spin wave excitations \cite{shapere_geometric_1990, pattanayak_gaussian_1994, sundaram_wave-packet_1999-1}, the full Lagrangian \Eq{eqn:Lag} can be simplified as
\begin{equation}
\label{eqn:Lag_ccm}
L = A^0_\mu(\varrho) \dot{X}_\mu - U_0(\varrho) +\rho \qty[\dot{\bx}\cdot (\bk-\ba) -A \bk^2 ],
\end{equation}
where the Einstein summation rule over repeated indices is assumed, and $\rho=\int \varrho \id \cV$ is the total spin wave intensity.
Here we use upper (lower) case to denote the texture (spin wave) quantities, which are connected via spatial integration, e.g.  $A=\int a \id \cV$.
For magnetic texture, $A_\mu^0(\varrho) = \int (1-\varrho)a^0_\mu \id \cV \simeq A^0_\mu - \rho a^0_\mu(\bx)$ is the vector potential for canonical coordinate $X_\mu$ with $a^0_\mu = \bm{\Lambda}_0 \cdot \partial_\mu \mb_0$,
$U_0(\varrho) = \int (1-2\varrho)u_0 \id \cV \simeq U_0 - 2\rho u_0(\bx)$ is the total texture energy, and both terms are slightly reduced due to the spin wave excitation $\varrho$.
For spin wave packets, $\bk = \bq+\ba$ is the canonical momentum, and $\bq$ is the central wave vector.
Similar to \Eq{eqn:Lag_ccm}, the Rayleigh function can also be rewritten in terms of the collective coordinates as: $R= (\alpha/2) \left[ \Gamma_{\mu\nu} \dot{X}_\mu \dot{X}_\nu + (\omega/  A)\rho\dot{x}_i^2 \right]$,
where $\Gamma_{\mu\nu}= \int (\partial_\mu \mb_0\cdot \partial_\nu \mb_0 )\id \cV$ is the dissipation for magnetic texture \cite{tretiakov_dynamics_2008}, and the second term is the dissipation for spin wave packet with $\omega$ the central frequency.

Invoking the Euler-Lagrangian rule for \Eq{eqn:Lag_ccm}, the dynamics of the magnetic texture and spin wave packet are governed by the following equations of motion:
\begin{subequations}
\label{eqn:eom_ccm}
\begin{align}
\label{eqn:eom_ccm_tx}
&E_{\mu}^0 +  B^0_{\mu\nu}\dot{X}_\nu - \alpha \Gamma_{\mu\nu} \dot{X}_\nu =   \rho b_{\mu i}\dot{x}_i \\
 \label{eqn:eom_ccm_sw}
&2e_i + b_{ij}\dot{x}_j + m_\ssf{sw}\ddot{x}_i+ \eta\dot{x}_i = - b_{\mu i} \dot{X}_\mu
\end{align}
\end{subequations}
where $m_\ssf{sw} = 1/2A$ is the normalized effective mass of the spin wave packet, and $\eta = \alpha\omega/A$ is the effective viscosity for spin wave packet.
Here $E^0_\mu(\varrho) = -\partial_\mu U_0(\varrho)$ and $B^0_{\mu\nu}(\varrho) = \partial_\mu A_\nu^0(\varrho)-\partial_\nu A_\mu^0(\varrho)$
are the effective electromagnetic fields in the parameter space spanned by the collective coordinates $\qty{X_\mu}$ for the magnetic texture.
Similarly $e_i= - \partial_i u_0$ and $b_{ij}= \partial_i a_j -\partial_j a_{i}$ are the effective electromagnetic fields for the spin wave packet located at $\bx$.
In addition, there are also effective magnetic fields across two parameter spaces of $\qty{X_\mu, x_i}$, \ie $b_{\mu i} = \partial_\mu a_i-\partial_i a_\mu$.
All magnetic fields  divide  into the contribution from magnetic topology and DMI respectively: $b= b^0+b^\ssf{D}$.
When written explicitly in terms of texture magnetization $\mb_0$, we have $b^0_{\mu \nu }= \mb_0 \cdot (\partial_\mu \mb_0 \times \partial_\nu \mb_0)$\cite{volovik_linear_1987-1}, which corresponds to the magnetic topology of the texture, and other fields $b^0_{ij}$, $b^0_{\mu i}$ follow simple index substitution $\mu,\nu\to i,j$.
The DMI induced magnetic fields read $b^\ssf{D}_{ij}=(D/2A)(\partial_i m^0_j-\partial_j m^0_i)$ and $b^\ssf{D}_{\mu i}=(D/2A)\partial_\mu m^0_i$.


As shown in \Eq{eqn:eom_ccm}, both the magnetic texture and the spin wave packet can be regarded as particle-like objects moving in their own parameter space $\{X_\mu\}$ and $\{x_i\}$ under influences of the corresponding effective electromagnetic fields. \cite{tretiakov_dynamics_2008, sundaram_wave-packet_1999-1} The texture is a massless particle, while the spin wave packet has an effective mass $m_\ssf{sw}$.
The magnetic texture and spin wave packet interact with each other via the mutual magnetic field $b^0_{\mu i}$ and $b^\ssf{D}_{\mu i}$ spanning across the parameter spaces $\qty{X_\mu, x_i}$.
Specifically, the right side of \Eq{eqn:eom_ccm_tx} and \Eq{eqn:eom_ccm_sw} refer to the  spin transfer torque exerted on magnetic texture,
and the texture-induced (electro-)motive force \cite{berger_possible_1986-2, barnes_generalization_2007-1, yang_universal_2009, guslienko_topological_2010-1, kovalev_thermomagnonic_2012-1} acting on spin wave respectively.
\emph{Spin wave skew scattering and side jump.} With the help of \Eq{eqn:eom_ccm}, we may analyze the scattering behavior between spin wave packet and the magnetic texture in a similar fashion as two particles collide. In particular, we consider a spin wave packet centered at $\bx(t)$ traveling upon a inhomogeneous magnetic texture with velocity $v_i$.
According to \Eq{eqn:eom_ccm_sw}, the spin wave packet, as a particle-like object, experiences an effective Lorentz force $f_j= v_i b_{ij}$ as it passes through the region of inhomogeneous magnetic texture.
Consequently a transverse velocity appears for the spin wave packet, similar to a  charged particle moving in real magnetic fields.
Assuming the transverse Lorentz force is small, and the consequent transverse velocity is much smaller than the original velocity $v_i$, the velocity correction due to the effective magnetic force is $\Delta v_j = m_\ssf{sw}^{-1}\int_{-\infty}^\infty dt f_j(t) $.
The force $f_j$ relies on the trajectory of the spin wave packet across the texture.
There are cases that this velocity correction vanishes: $\Delta v_j = 0$, but the position correction is always non-zero: $\Delta x_j = m_\ssf{sw}^{-1}\int_{-\infty}^\infty dt \int_{-\infty}^t dt' f_j(t')$. This transverse velocity and position corrections define the skew scattering and the side jump for a spin wave packet traveling across the texture.

\begin{figure}[t]
\centering
 \includegraphics[width=0.4\textwidth ]{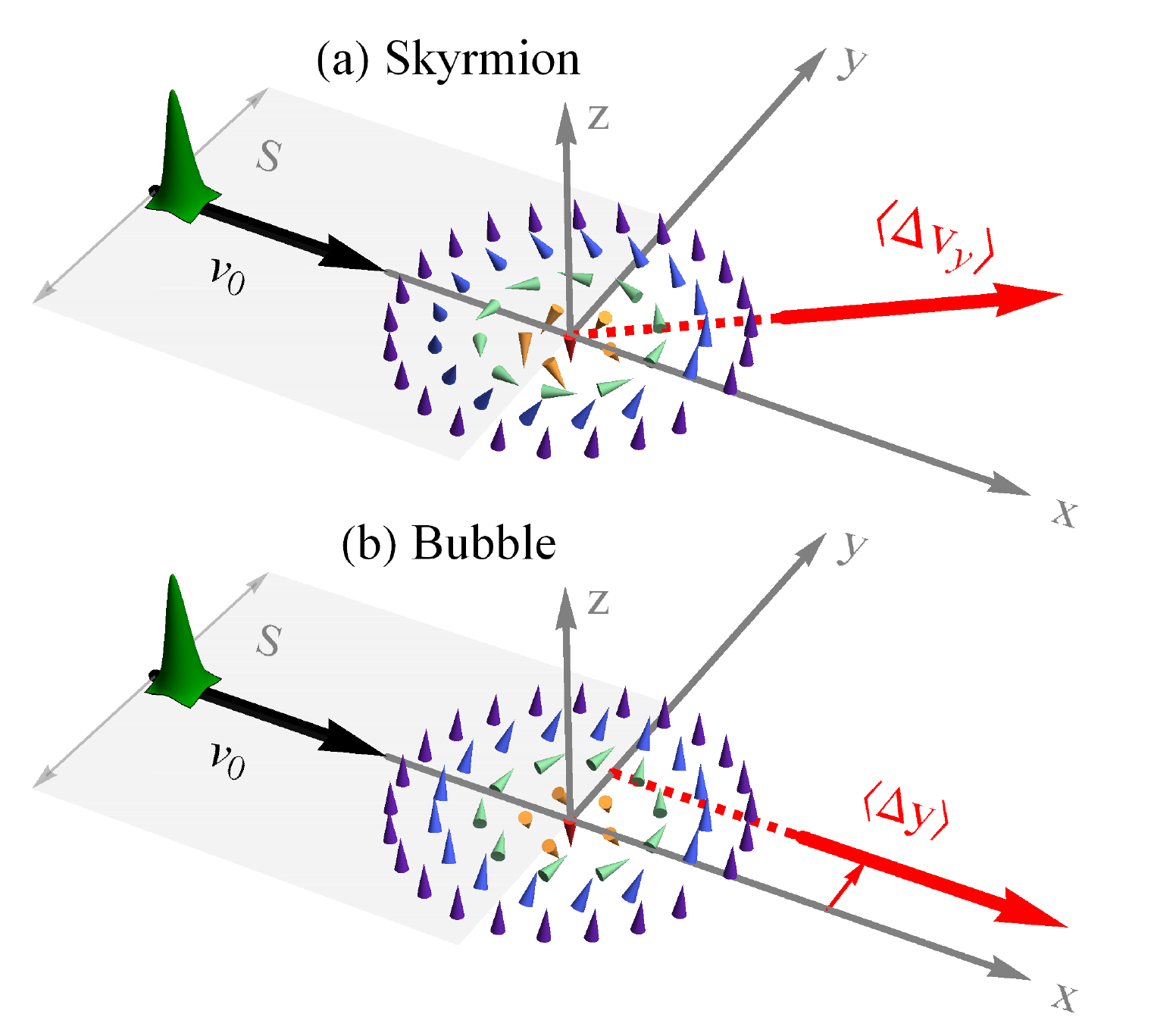}
\caption{Schematics of (a) spin wave skew scattering across magnetic Skyrmion and (b) spin wave side jump across magnetic bubble.
The spin wave beam incidents on the Skyrmion/bubble in the $x$ direction with velocity $v_0$ and cross section $S$, and the scattering by magnetic texture leads to:
(a) the out-going beam acquires an averaged velocity $\langle\Delta v_y\rangle $, and forms an angle with incident beam;
(b) the out-going beam is parallel to incident beam, but takes an averaged  shift $\langle \Delta y\rangle$.
}
\label{fig:sch}
\end{figure}

With the above knowledge about a single spin wave packet, we now consider a spin wave beam with cross section $S$ passing through a magnetic texture with velocity $v_0$. This spin wave beam can be regarded as a collection of many spin wave packets, each of which passing through the texture via a different trajectory. Assuming that the cross section of the spin wave beam covers the whole texture,
then the Lorentz force is averaged over the whole texture region according to \Eq{eqn:eom_ccm_sw}, and the average transverse velocity is calculated as
\begin{equation}
\label{eqn:Delta_vel_sw}
\expval{\Delta v_j}
\simeq \frac{ 2A}{ S} B^0_{ij} + \frac{ D}{S} \qty(\pdv{ M^0_i}{X_j}-\pdv{ M^0_j}{X_i}).
\end{equation}
Besides a contribution from the magnetic topology induced field: $B^0_{ij}=\int b^0_{ij} \id \cV=\int \mb_0 \cdot (\partial_i \mb_0\times \partial_j \mb_0)\id \cV$, the skew scattering is also correlated to the variation of the total texture magnetization $M^0_{i/j}=\int m^0_{i/j}\id \cV$ on texture position $X_{j/i}$.
which has already been demonstrated by spin wave crossing a chiral domain wall \cite{yu_magnetic_2016}.
The skew scattering $\expval{\Delta v_j}$ is independent of incident velocity $v_0$ because of the cancellation between the Lorentz force ($\propto v_0$) and the traveling time ($\propto 1/v_0$) across the magnetic texture.
Similarly, the average side jump is calculated as
\begin{equation}
\label{eqn:Delta_pos_sw}
\expval{\Delta x_j}
\simeq \frac{D}{S} \frac{M^0_j}{v_0},
\end{equation}
and it is related to the total magnetization $M^0_j$ and the strength of DMI. The side jump $\expval{\Delta x_j}$ is reciprocal to $v_0$ because it takes time for the the positional shift to accumulate.

Both skew scattering and side jump are affected by the texture dynamics, as seen in the right side of \Eq{eqn:eom_ccm_sw}.
Besides the Lorentz force, the spin wave packet is also subject to the electrostatic force $f_i=-2e_i$, and a longitudinal position correction (or forward jump) arises.
Similar to the side jump in \Eq{eqn:Delta_pos_sw}, the averaged forward jump of a spin wave beam is $\langle \Delta x_i \rangle \simeq (2A/S) (U_0/v^2_0)$, which is proportional to the texture energy $U_0$.

\begin{figure}[t]
\centering
\includegraphics[width=0.5\textwidth]{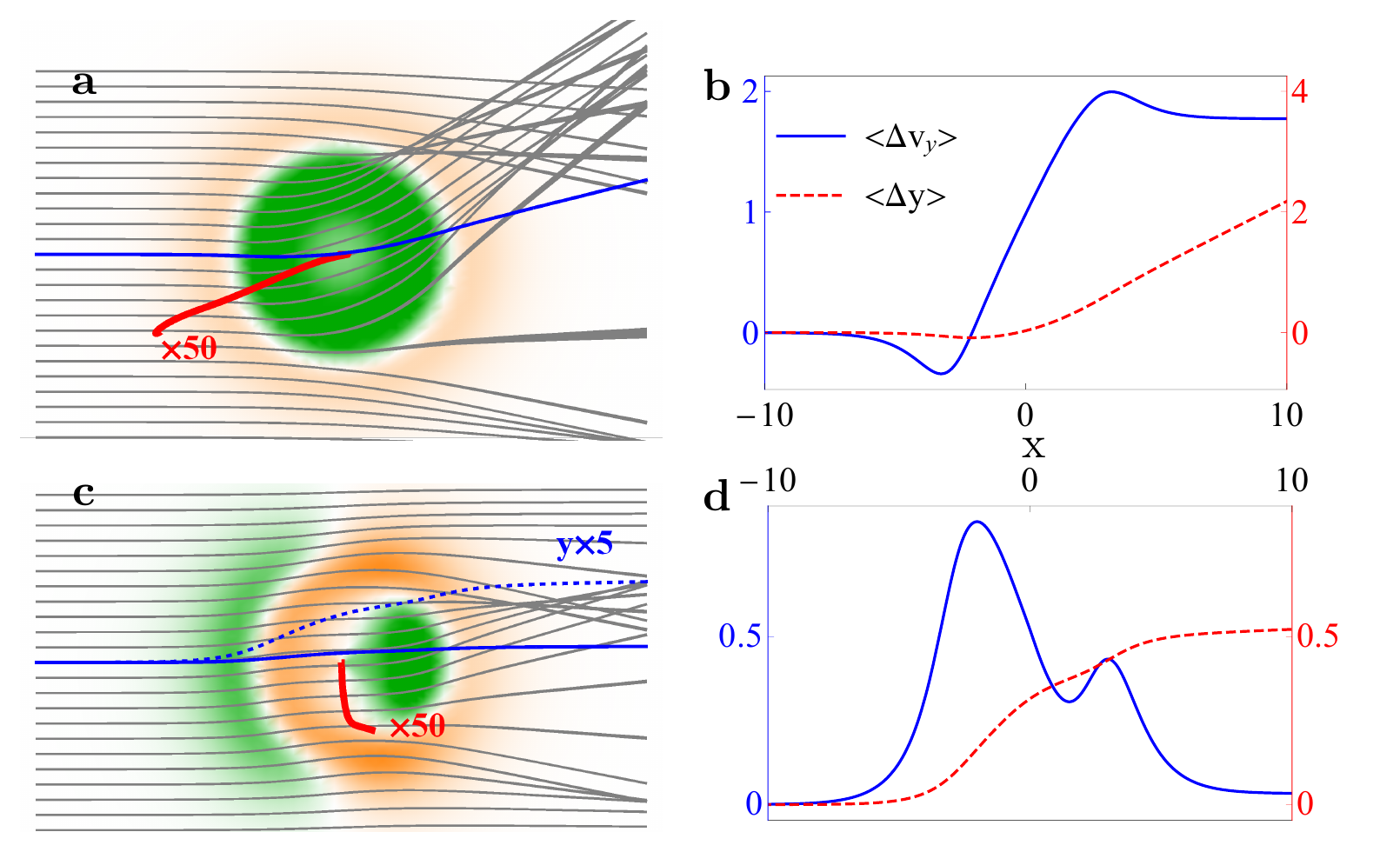}
\caption{Skew scattering and side jump across (a)(b) Skyrmion and (c)(d) bubble in numerical calculations. (a)(c) depict the spin wave trajectories in real space, where each gray line represents trajectory of one spin wave packet. The blue line represent the averaged trajectory of $625$ spin wave packets, and the red line is for texture trajectory (exaggerated by $50$ times). The magnetic field $b$ is encoded in the background color, where green/orange color are for positive/negative value respectively. (b)(d) plot the averaged velocity $\expval{\Delta v_y}$ and position $\expval{\Delta y}$ as function of position $x$. In calculations, following parameters are used: the incident velocity $v_0=10$, spin wave intensity $\rho=0.01$, the bubble mass $m_\ssf{bub}=10$, and magnetic constants $A=1$, $K=1$, $D=1.2$.}
\label{fig:num}
\end{figure}

\emph{Spin wave scattering across magnetic Skyrmion and bubble.} We now discuss the specific magnetic textures of
magnetic Skyrmion and magnetic bubble with their typical magnetic profiles depicted in Fig. \ref{fig:sch}. A magnetic topological charge \cite{nagaosa_topological_2013, koshibae_berry_2016} for such types of texture in a 2-dimensional plane can be defined as: $Q=(1/4\pi)\int \mb_0 \cdot (\partial_x \mb_0\times \partial_y \mb_0)\dd{x}\dd{y}$.
The Skyrmion has non-zero topological charge $Q \neq 0$, while bubble is the topologically trivial with $Q= 0$.
Magnetic Skyrmion and bubble can be hosted in the same magnetic film under  slightly different external magnetic fields  \cite{phatak_nanoscale_2016, loudon_images_2019}.
The texture-induced effective magnetic field is $B^0=B^0_{xy}= \int \mb_0 \cdot (\partial_x \mb_0\times \partial_y \mb_0)\dd{x}\dd{y} = 4\pi Q$, which vanishes for magnetic bubble, respectively.
From symmetry analysis, the in-plane components of the total magnetization of Skyrmion is zero: $M^{x, y}_0 = 0$, but for bubble is $M_0^{x, y} \neq 0$. In addition, the total magnetization of Skyrmion or bubble is independent of its location, i.e. $\partial M^0_x/\partial Y = \partial M^0_y/\partial X=0$. Based on these characters of Skyrmion and bubble, we may infer from \Eq{eqn:Delta_vel_sw} that the spin wave beam would experience skew scattering across the Skyrmion due to the non-zero $B^0$, as shown in \Figure{fig:sch}(a). The spin wave beam would have no skew scattering when scattering across a bubble, instead, a finite side jump and forward jump would appear according to \Eq{eqn:Delta_pos_sw}, as shown in Fig. \ref{fig:sch}(b).


The minimal set of collective coordinates for both Skyrmion and bubble are their central positions $\bX=(X, Y)$, then \Eq{eqn:eom_ccm} reduces to equations in real space:
\begin{subequations}
\label{eqn:eom_bubble}
\begin{align}
\label{eqn:eom_bubble_tx}
 \dot{\bX}\times\bB^0 &= -  \rho  \dot{\bx} \times \bb,   \\
\label{eqn:eom_bubble_sw}
m_\ssf{sw} \ddot{\bx} + \dot{\bx} \times (\bb^0+\bb^\ssf{D}) &= \dot{\bX}\times \bb,
\end{align}
\end{subequations}
where the electric field and small corrections to texture fields are neglected, and the DMI induced magnetic field are symmetrized for simplicity.
The effective magnetic fields $\bb=(b^0+b^\ssf{D})\hbz$ is perpendicular to the film with $b^0= \mb_0\cdot (\partial_x \mb_0 \times \partial_y \mb_0)$ and $b^\ssf{D}=(D/2A) (\nabla\times \mb_0)\cdot\hbz$. The spatial distribution for field $b$ are shown in Fig. \ref{fig:num}(a)(c), where the field is rotationally symmetric for magnetic Skyrmion, but asymmetric for the magnetic bubble.

\begin{figure}[t]
\centering
\includegraphics[width=0.49\textwidth]{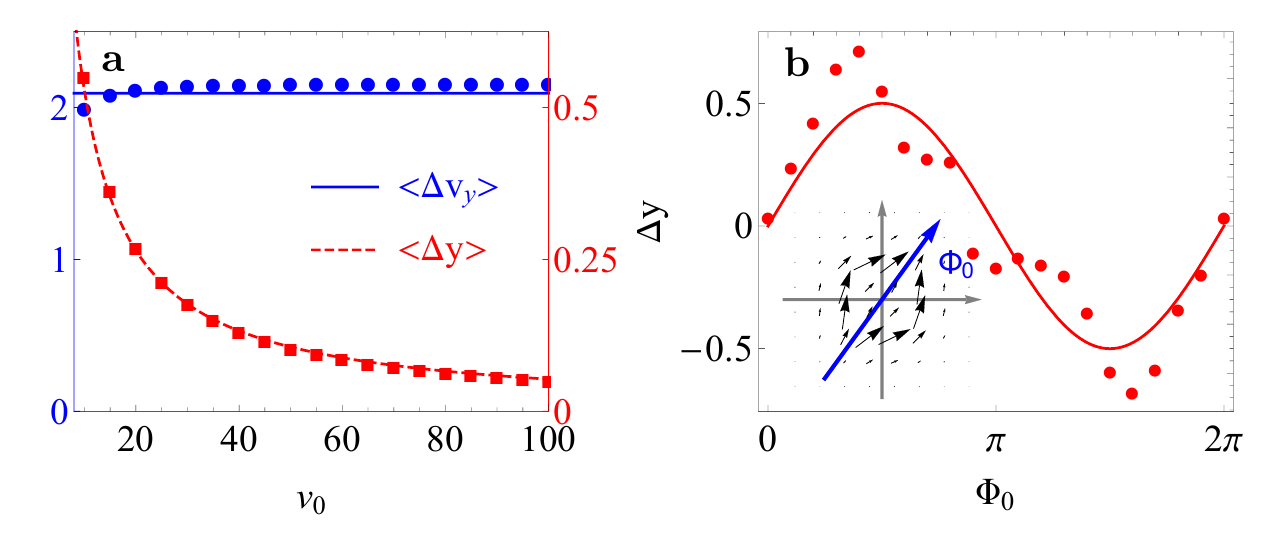}
\caption{
(a) Skew scattering and side jump as function of incident velocity $v_0$.
The skew scattering $\langle \Delta v_y \rangle$ (in blue) is for the Skyrmion case, and the side jump $\langle \Delta y\rangle$ (in red) is for the bubble case.
The dots are data extracted from numerical calculations based on \Eq{eqn:eom_bubble}, and the solid lines follows \Eq{eqn:Delta_vel_sw} and \Eq{eqn:Delta_pos_sw}.
(b) The side jump $\langle \Delta y \rangle$ as function of the angle $\Phi_0$, with the definition of $\Phi_0$ depicted in inset.
}
\label{fig:num_data}
\end{figure}

For a spin wave packet interacting with magnetic Skyrmion (bubble), its trajectory can be calculated from \Eq{eqn:eom_bubble} as one classical particle (spin wave packet) penetrating another one (Skyrmion or bubble).
In Fig. \ref{fig:num}(a)(c), $625$ spin wave packets in $25$ row and $25$ column are prepared and pass through the Skyrmion/bubble.
As demonstrated, the exact trajectory of a spin wave packet depends on the impact parameter (the vertical shift from the bubble core).
In general, those wave packets with smaller impact parameter are more strongly deflected due to the larger field near the core.
The averaged trajectory is bent upward (skew scattered) across the magnetic Skyrmion, and is shifted upward (side jumped) across the magnetic bubble, as demonstrated in \Figure{fig:num}(a)(c).

The Skyrmion or bubble experience the reaction force from the spin wave scattering and its motion can be solved via \Eq{eqn:eom_bubble_tx}.
The Skyrmion is found to be pulled by the spin wave and moves to the lower-left, because Skyrmion is also deflected by its own field $\bB^0$;
while the magnetic bubble moves to lower-right, because its motion is mainly controlled by its inertia (characterized as $m_\ssf{bub}$) induced by distortion \cite{makhfudz_inertia_2012, schutte_inertia_2014-1, koshibae_berry_2016}, since the topological protection is absent $B^0=Q=0$.
The motion of Skyrmion/bubble in turn slightly modifies the trajectories of subsequently passing spin wave packets, highlighting their mutual interaction \cite{schutte_Magnonskyrmion_2014,iwasaki_Theory_2014}.

The skew scattering and side jump are further indicated by the averaged transverse velocity $\expval{\Delta v_y}$ and position-shift $\expval{\Delta y}$ as shown in Fig. \ref{fig:num}(b)(d). For Skyrmion case, the spin wave beam finally acquires a finite transverse velocity $\expval{\Delta v_y}$ so that its transverse position $\expval{\Delta y}$ continuously increases even after spin wave moves away from the Skyrmion.
In contrast, for magnetic bubble case, the transverse velocity $\expval{\Delta v_y}$ is finite only when inside the bubble, and vanishes when the spin wave moves away from the bubble, but its accumulation leads to the a finite transversal shift $\expval{\Delta y}$.
The corrections $\expval{\Delta v_y} $ in skew scattering and $\expval{\Delta y}$ in side jump for different incident velocities are further investigated in Fig. \ref{fig:num_data}(a).
The almost constant $\expval{\Delta v_y}$ and the reciprocally decreasing $\expval{\Delta y}$ agree well with \Eq{eqn:Delta_vel_sw} and \Eq{eqn:Delta_pos_sw}.
The Skyrmion is rotationally invariant, but the magnetization of magnetic bubble depends on the angle $\Phi_0$ with $x$-axis as defined in Fig. \ref{fig:num_data}(b) inset.
As a result, when the magnetic bubble rotates in $x$-$y$ plane, its total magnetization $M_0^y$ varies sinusoidally with $\sin\Phi_0$, and the side jump $\expval{\Delta y}$ varies accordingly in Fig. \ref{fig:num_data}(b).

\emph{Discussions.} Skew scattering and side jump are two basic scattering patterns by impurity \cite{berger_side-jump_1970}, and the two basic extrinsic mechanisms for anomalous Hall effect in electronic system \cite{takahashi_spin_2008, sinova_spin_2015}.
Here we show that besides the magnetic topology, the DMI also plays important role in spin wave skew scattering and side jump,
similar to the spin orbital coupling in its electronic counterpart.

In this work, the collective coordinate model is focused on the ferromagnetic environment, but it naturally extends to antiferromagnetic system.
The magnetic texture in antiferromagnets becomes massive instead of gyroscopic \cite{tveten_staggered_2013-1}.
In addition, a full set of polarization modes are hosted by antiferromagnets \cite{lan_antiferromagnetic_2017-1}, which can be formulated by the non-Abelian wave packet theory \cite{culcer_coherent_2005, daniels_nonabelian_2018-1}.
The extension to other cases such as the system with the hard-axis anisotropy, or ferrimagnetic environment also follow similar procedures as outlined above.

\emph{Conclusions.} In conclusion, by simplified trajectory analysis in collective coordinates, we show that spin wave may experience both skew scattering and side jump across magnetic texture, depending on the magnetic topology and the total magnetization of the magnetic texture.
The concept of skew scattering and side jump allows us to understand the interaction between spin wave and magnetic texture in a more intuitive way.

\end{document}